\documentclass[prd,aps,preprintnumbers,nofootinbib,superscriptaddress,notitlepage,floatfix,11pt]{revtex4}
\usepackage{stmaryrd}
\usepackage{epsfig}
\usepackage{amsfonts}
\usepackage{amsmath}
\usepackage{graphicx}
\usepackage{color}
\usepackage{amssymb,amsmath,psfrag,slashed,graphicx}
\usepackage[normalem]{ulem}

\def \nn {\nonumber}

\allowdisplaybreaks

\begin{document}
\title{Nuclear effects in neutrino-nucleus DIS and probe for short range correlation}

\affiliation{School of Physics and Microelectronics, Zhengzhou University, Zhengzhou, Henan 450001, China}
\affiliation{INPAC, Key Laboratory for Particle Astrophysics and Cosmology (MOE),
Shanghai Key Laboratory for Particle Physics and Cosmology,
School of Physics and Astronomy, Shanghai Jiao Tong University, Shanghai 200240, China}
\affiliation{School of Physics and Optoelectronic Engineering, Shandong University of Technology, Zibo, Shandong 255000, China}

\author{Fei Huang~\footnote{Corresponding author. fhuang@sjtu.edu.cn}}
\affiliation{INPAC, Key Laboratory for Particle Astrophysics and Cosmology (MOE),
Shanghai Key Laboratory for Particle Physics and Cosmology,
School of Physics and Astronomy, Shanghai Jiao Tong University, Shanghai 200240, China}

\author{Ji Xu~\footnote{Corresponding author. xuji\_phy@zzu.edu.cn}}
\affiliation{School of Physics and Microelectronics, Zhengzhou University, Zhengzhou, Henan 450001, China}

\author{Xing-Hua Yang~\footnote{Corresponding author. yangxinghua@sdut.edu.cn}}
\affiliation{School of Physics and Optoelectronic Engineering, Shandong University of Technology, Zibo, Shandong 255000, China}

\begin{abstract}
We investigate charged-current neutrino-nucleus deep inelastic scattering with particular interests in relationship of short range correlation (SRC) and EMC effect. The structure functions $F^A_2(x,Q^2)$, $x F^A_3(x,Q^2)$ and ratios of differential cross sections are presented where the nuclei $A$ are chosen to be carbon, iron and lead. We propose a kind of universal modification functions which would provide a nontrivial test of SRC universality on the platform of neutrino-nucleus scattering and improve our understanding of nucleon structure substantially.
\end{abstract}

\maketitle

\section{Introduction}
\label{sec:Introduction}

As elementary particles in standard model with no electric charge, neutrinos interact only via the weak interaction which can be categorized by neutrino energies. Elastic scattering has been used to measure the nuclear charge distribution. Quasi elastic scattering (QES) can serve as a normalisation cross section when experimentalists measuring cross section ratios of neutrino scattering \cite{Zyla:2020zbs}. Resonance pion production (RES) is of critical importance to the current and future neutrino oscillation experiments \cite{Giganti:2017fhf}. Neutrino-nucleus deep inelastic scattering (DIS) is sensitive to the nuclear quark distributions \cite{Hen:2016kwk}. As the energy of neutrinos increases, the DIS process becomes dominant in neutrino-nucleus interactions.

One cannot determine the individual parton distribution functions (PDFs) through electron DIS experiments alone, information on the flavor decomposition of PDFs can be obtained by neutrino DIS experiments which are crucial for global fits of PDFs, especially for the strange quark distribution. A good knowledge of neutrino DIS is also a key factor for long baseline experiments aiming at measuring the mixing angle and CP-violating phase. In this process, a neutrino scatters with a quark inside a nucleon ``deeply'' (high momentum transfer $Q^2$) at high energy which destroys the nucleon. Because of the weak nature of neutrino interactions, the use of heavy nuclear targets is unavoidable, this brings complications since correlations between nucleons, dubbed as nuclear effects, can affect observables in neutrino interaction. Nuclear effects are still not well understood in neutrino physics and there have been renaissance interests to investigate the nuclear effects in neutrino DIS process with the following reasons. Firstly, the data of current neutrino DIS measurements are valuable because $F_3$ structure function can directly probe valence quark distributions. But the data contain non-negligible nuclear effects since heavy nuclear targets such as iron and plumbum are used. Secondly, there is an issue of NuTeV anomaly which opens the question of factorization assumption in the nuclear environment, whether the neutrino DIS data could be combined with the charged-lepton DIS data to get a better nuclear PDFs (NPDFs) \cite{Tzanov:2005kr,Schienbein:2007fs,Schienbein:2009kk,Kovarik:2010uv,Paukkunen:2010hb}. This issue has attracted a lot of interests, Ref. \cite{Cloet:2009qs} pointed out that this anomaly might be due to differences between the proton and the neutron caused by mean-field effects and a nucleon-nucleon short-range correlation (SRC) explanation is also presented \cite{Schmookler:2019nvf}. Thirdly, nuclear effects are important in analyzation of high energy neutrino interacting with Earth matter, the total cross section is dominated by DIS off the partons from matter nucleons, therefore the nuclear effects have significant impact on the predictions for the neutrino attenuation rates \cite{Garcia:2020jwr,Bertone:2018dse,Wang:2019dal,Xing:2019owb,Li:2016txk,Xing:2000ik}.

With such an importance, neutrino DIS and nuclear effects are explored worldwide, e.g., NuTeV \cite{Tzanov:2005kr}, CHORUS \cite{Onengut:2005kv}, MINOS \cite{Adamson:2009ju} and MINERvA \cite{Mousseau:2016snl} in recent years. They have analyzed neutrino DIS off different nuclear targets to measure structure functions, differential and total cross sections. Reconciled with the charged-lepton DIS data and Drell-Yan (DY) dilepton production data etc., the NPDFs can be extracted through global fitting \cite{Eskola:2012rg,Paukkunen:2014nqa}. One can resort to parametrizations of nCTEQ \cite{Kovarik:2015cma} and EPPS16 \cite{Eskola:2016oht} for more details.

The original idea of having nuclear effects in PDFs was driven by data in DIS measurements performed by the European Muon Collaboration (EMC) \cite{Aubert:1983xm}. The initial expectation was that physics at GeV scale would be insensitive to the nuclear binding effects which are typically on the order of several MeV scale. However, the collaboration discovered the per-nucleon deep inelastic structure function in iron is smaller than that of deuterium in the region $0.3<x<0.7$, here $x$ is the Bjorken variable. This phenomenon is known as EMC effect and has been observed for a wide range of nuclei \cite{Arneodo:1988aa,Arneodo:1989sy,Allasia:1990nt,Gomez:1993ri,Seely:2009gt}. EMC effect indicates that quark PDFs in nucleon are modified, breaking down the scale separation between nucleon structure and nuclear structure. Although the understanding of how the quark-gluon structure of a nucleon is modified by the surrounding nucleons has been brought to a whole new level, one should note that there is still no consensus as to the underlying dynamics that drives this effect even after more than three decades.

Currently, one of the leading approaches for describing the EMC effect is: nucleons bound in nuclei are unmodified, same as ``free'' nucleons most of the time, but are modified substantially when they fluctuate into SRC pairs. The SRC describes the probability that two nucleons are close in coordinate space, as a result of nontrivial nucleon-nucleon interactions in nucleus. The connection between SRC and EMC effects has been extensively investigated in nuclear structure function measurements \cite{Egiyan:2005hs,Hen:2012fm,Hen:2014nza,Duer:2018sby}. A linear relation between the magnitude of the EMC effect measured in electron DIS at $0.3<x<0.7$ and SRC scale factor obtained from electron inelastic scattering at $x>1.5$ was proposed \cite{Weinstein:2010rt}, this striking linear relation suggests that the EMC effect and SRC correlation both stem from the same underlying physics such as high local density and high momentum nucleons in nuclei. This relation, if finally established, shall provide a unique method to study nuclear structure physics, see some of recent developments in Refs. \cite{Chen:2016bde,Lynn:2019vwp,Xu:2019wso,Hatta:2019ocp}. One of the key aspects of SRC is the universality, where the partonic structure from the correlated nucleon-nucleon pair is same for all kinds of nuclei, thus a universal modification function can be deduced. This function will be useful for testing QCD symmetry breaking mechanisms and for distinguishing nuclear physics effects from beyond the standard model effects in neutrino scattering experiments.

The robust linear correlation between the strength of the EMC and the SRC scale factor in nuclei indicates possible modifications of the quark PDFs occur in nucleons which are inside SRC pairs. As illustrated above, neutrino-nucleus DIS process is another ideal platform for testing nucleon structures and SRC interpretation of the EMC effect, it is sensitive to the quark distributions, especially for $u$ and $d$ quark (antiquark), they contain crucial information on nuclear effects which should be distinguished from beyond the standard model effects \cite{Peskin:1995ev,Grover:2018ggi,Hen:2013oha,Li:2018wut,Si:2017pdo,Huang:2017yso,Yang:2020sos}. These can be studied by measuring three structure functions $F_1(x,Q^2)$, $F_2(x,Q^2)$ and $F_3(x,Q^2)$. In this work, we will study charged-current neutrino-nucleus deep inelastic scattering where the nuclei $A$ are chosen to be $\rm{^{12}C}$, $\rm{^{56}Fe}$ and $\rm{^{208}Pb}$. The structure functions $F^A_2(x,Q^2)$ and $x F^A_3(x,Q^2)$ are calculated with consideration of NPDFs by using EPPS16 parametrization \cite{Eskola:2016oht}, our free nucleon baseline is CT14NLO \cite{Dulat:2015mca}. The ratios of structure functions for nucleus and free nucleon as well as ratios of differential cross sections $d^2\sigma/dxdy$ are also presented for a better view of EMC effect. A kind of universal modification functions were deduced in neutrino-nucleus DIS, these functions would provide a nontrivial test of SRC universality on the platform of neutrino-nucleus scattering.

The rest of this paper is organized as follows. The formalism and results of structure functions in neutrino-nucleus DIS are presented in In Sec.~\ref{structurefunc}. Differential cross sections $(d^2\sigma/dxdy)^{\nu A}$ and $(d^2\sigma/dxdy)^{\bar\nu A}$ are analysed in Sec.~\ref{diffcrosssection}. Sec.~\ref{universality} is devoted to the discussion of universal modification functions. Finally, we summarize our paper and comment on the future developments in Sec.~\ref{conclusion}.

\section{Formalism for structure functions in neutrino-nucleus DIS}
\label{structurefunc}
\begin{figure}
\includegraphics[width=0.4\columnwidth]{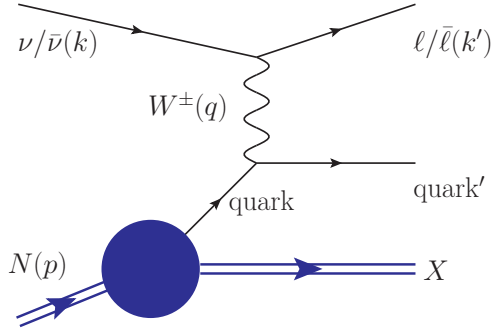}
\caption{Neutrino charged-current DIS process.}
\label{CC_DIS}
\end{figure}

Neutrino charged-current DIS depicts the process in which a neutrino scatters off a quark in the nucleon via the exchange of $W^{\pm}$ boson whose momentum is $q$, producing a corresponding lepton and hadron in the final state, which can be written as
\begin{eqnarray}
  \nu_\ell(k)+N(p)\to \ell(k')+X(p') \,, \qquad \bar\nu_\ell(k)+N(p)\to \bar\ell(k')+X(p') \,.
\end{eqnarray}
Here $k$ is the momentum of incoming neutrino (antineutrino) and $k'$ indicates the momentum of outgoing lepton (antilepton). The momentum of  initial nucleon and produced hadron are $p$ and $p'$ respectively. These processes can be analyzed by the following Lorentz invariants: the Bjorken scaling variable $x\equiv Q^2/(2p\cdot q)$, the inelasticity $y\equiv(2p\cdot k)/(2p\cdot q)$, the negative squared four momentum transfer $Q^2\equiv -q^2$. A schematic diagram of the neutrino charged-current DIS process is shown in Fig.\ref{CC_DIS}.

The amplitude for neutrino charged-current interactions with the nucleon is described by the matrix element
\begin{eqnarray}
  \mathcal{M}=\frac{G_{F} / \sqrt{2}}{1+Q^{2} / M_{W}^{2}} \bar{u}\left(k^{\prime}\right) \gamma^{\mu}\left(1-\gamma_{5}\right) u(k)\left\langle X\left|J_{\mu}^{C C}(0)\right| p\right\rangle \,,
\end{eqnarray}
where $G_F$ is the Fermi coupling constant and $J_{\mu}^{C C}(0)$ is the weak charged-current of the nucleon. The leptonic part can be calculated perturbatively, the hadronic part becomes the hadron tensor $W^{CC}_{\mu\nu}$, which is then expressed by three structure functions in the expression of neutrino double differential cross sections \cite{Hirai:2009mq}
\begin{eqnarray}\label{diff_for_nuandbarnu1}
  \left(\frac{d \sigma}{d x d y}\right)^{\nu A}&=&\frac{G_{F}^{2} M_{N} E_\nu}{\pi\left(1+Q^{2} / M_{W}^{2}\right)^{2}}\left[F_{1}^{\nu A}(x,Q^2) x y^{2}+F_{2}^{\nu A}(x,Q^2)\left(1-y\right) + F_{3}^{\nu A}(x,Q^2) x y\left(1-\frac{y}{2}\right)\right] \,,\nn\\
  \left(\frac{d \sigma}{d x d y}\right)^{\bar\nu A}&=&\frac{G_{F}^{2} M_{N} E_{\bar\nu}}{\pi\left(1+Q^{2} / M_{W}^{2}\right)^{2}}\left[F_{1}^{\bar\nu A}(x,Q^2) x y^{2}+F_{2}^{\bar\nu A}(x,Q^2)\left(1-y\right) - F_{3}^{\bar\nu A}(x,Q^2) x y\left(1-\frac{y}{2}\right)\right] \,.\nn\\
\end{eqnarray}
Here $E_{\nu(\bar\nu)}$ is the energy of incoming neutrino (antineutrino), $M_N$ is the nucleon mass. The expressions above can be further simplified by using Callan-Gross relation $F_2^{\nu A}=2x F_1^{\nu A}, F_2^{\bar\nu A}=2x F_1^{\bar\nu A}$ \cite{Callan:1969uq},
\begin{eqnarray}\label{diff_for_nuandbarnu2}
\left(\frac{d \sigma}{d x d y}\right)^{\nu A}&=&\frac{G_{F}^{2} M_{N} E_\nu}{\pi\left(1+Q^{2} / M_{W}^{2}\right)^{2}}\left[\left(1-y+\frac{y^{2}}{2}\right) F_{2}^{\nu A }+x y\left(1-\frac{y}{2}\right) F_{3}^{\nu A}\right] \,,\nn\\
\left(\frac{d \sigma}{d x d y}\right)^{\bar{\nu} A}&=&\frac{G_{F}^{2} M_{N} E_{\bar\nu}}{\pi\left(1+Q^{2} / M_{W}^{2}\right)^{2}}\left[\left(1-y+\frac{y^{2}}{2}\right) F_{2}^{\bar \nu A}-x y\left(1-\frac{y}{2}\right) F_{3}^{\bar \nu A}\right] \,,
\end{eqnarray}
with
\begin{eqnarray}
  \left\{\begin{array}{ll}
  F_2^{\nu A}=\frac{Z}{A} F_2^{\nu p/A} + \frac{A-Z}{A} F_{2}^{\nu n /A} \,,\\
  F_2^{\bar \nu A}=\frac{Z}{A} F_2^{\bar\nu p/A} + \frac{A-Z}{A} F_{2}^{\bar\nu n /A} \,.
\end{array} \right. \nn
\end{eqnarray}
Here the explicit dependence on $x$ and $Q^2$ is omitted for brevity. We would discuss what kind of combinations of NPDFs can be constrained by the neutrino-nucleus DIS data, in parton model, the structure functions for a nucleus with atomic mass $A$ and atomic number $Z$ can be expressed in terms of NPDFs (also suppressing the dependence on $x$ and $Q^2$):
\begin{eqnarray}\label{calcstrucuturefuncs1}
  F_2^A &=& F_2^{\nu A} + F_2^{\bar\nu A} \nn\\
  &=& \frac{Z}{A}\left(F_2^{\nu p/A}+F_2^{\bar\nu p/A}\right) + \frac{A-Z}{A}\left(F_2^{\nu n/A}+F_2^{\bar\nu n/A}\right) \nn\\
  &=& 2x \frac{Z}{A} \left( f_u^{p/A}+f_{\bar u}^{p/A} +f_d^{p/A}+f_{\bar d}^{p/A} \right) + 2x\frac{A-Z}{A}\left( f_u^{n/A}+f_{\bar u}^{n/A} +f_d^{n/A}+f_{\bar d}^{n/A} \right) \,,
\end{eqnarray}
where $f_i^{p/A}$ represents NPDFs of parton $i$ in proton bound in nucleus $A$ and $f_i^{n/A}$ corresponds to neutron. The NPDFs of strange and charm quark are not considered here since the contributions are very small in the EMC and kinematic regions where this work focuses on.

\begin{figure}[htbp]
\centering
\includegraphics[width=0.45\columnwidth]{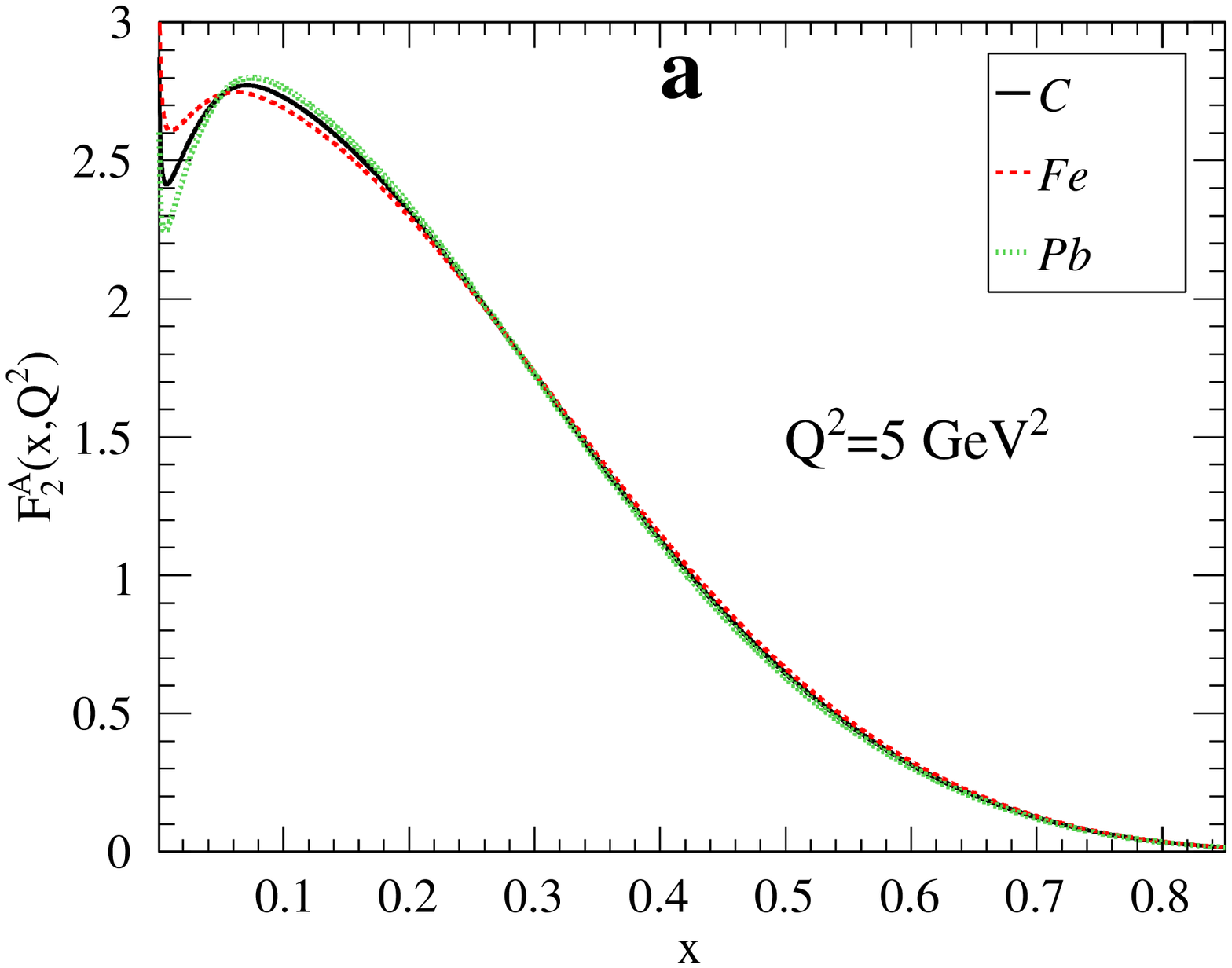}
\centering
\includegraphics[width=0.45\columnwidth]{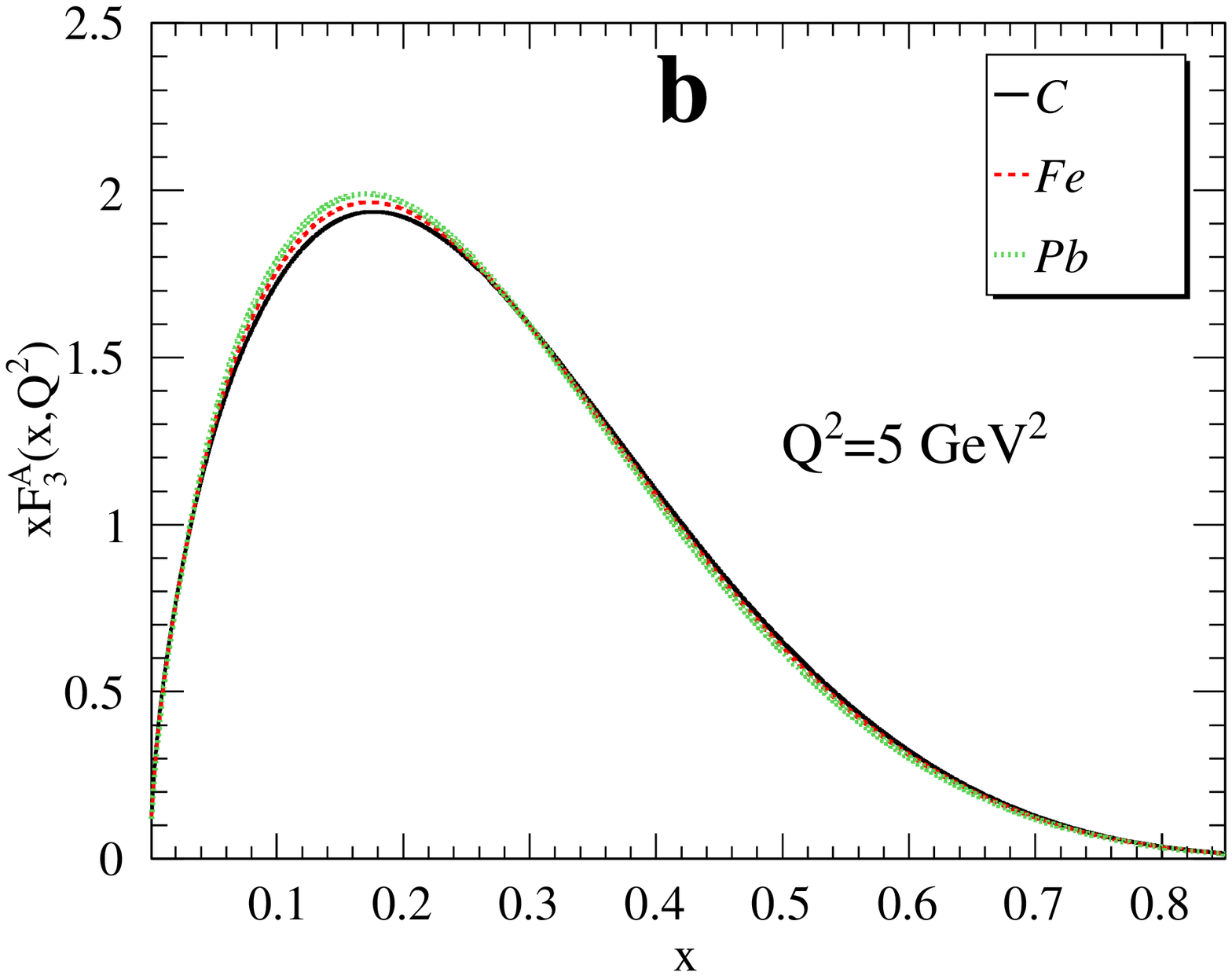}
\centering
\caption{Structure functions with EPPS16 parametrization of nuclear corrections for $\rm{C}$, $\rm{Fe}$ and $\rm{Pb}$ as functions of $x$ with $Q^2=5~\textrm{GeV}^2$. $F_2(x,Q^2)$ and $xF_3(x,Q^2)$ are presented in \textbf{(a)} and \textbf{(b)} respectively.}
\label{structure_functions}
\end{figure}

\begin{figure}[htbp]
\centering
\includegraphics[width=0.45\columnwidth]{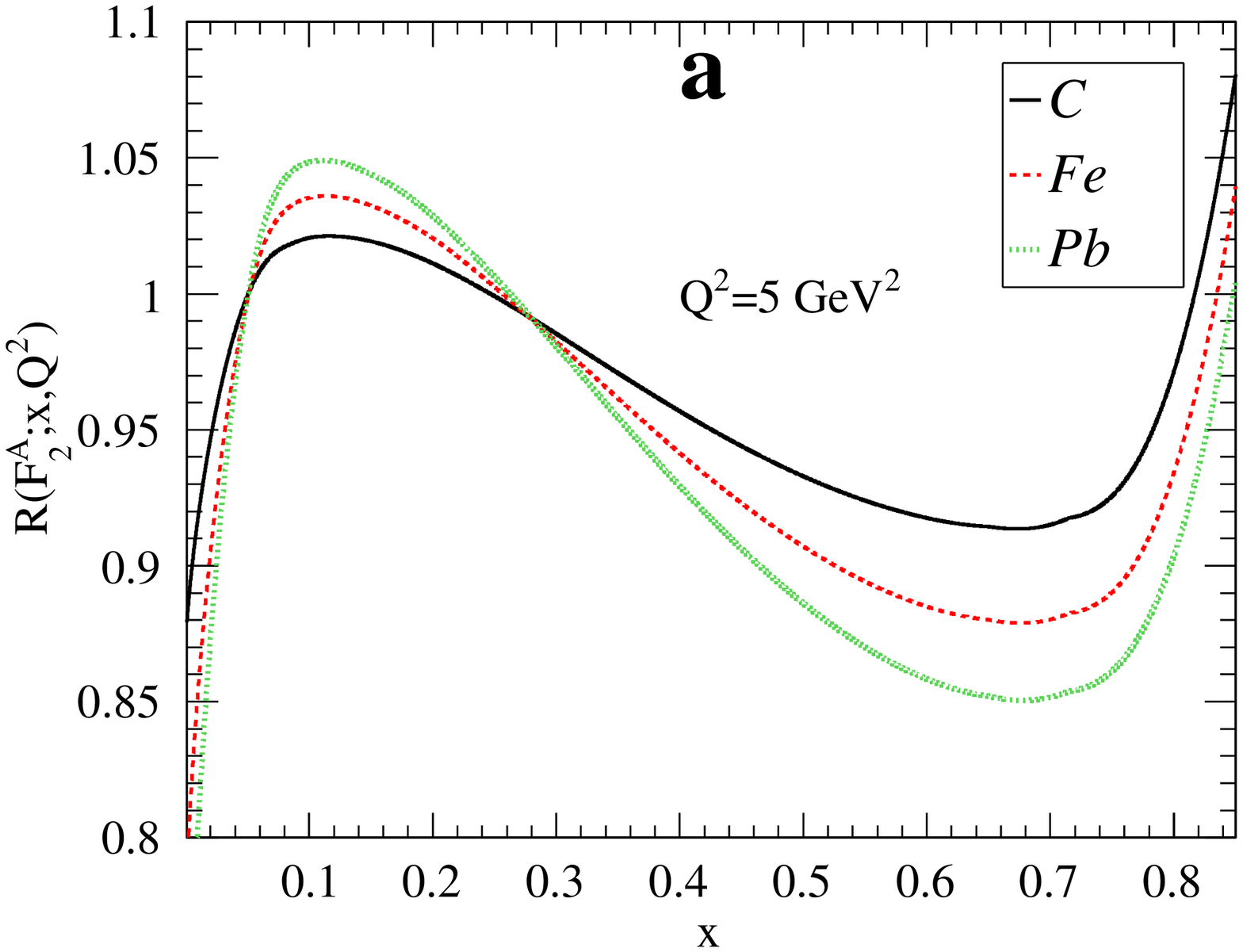}
\centering
\includegraphics[width=0.45\columnwidth]{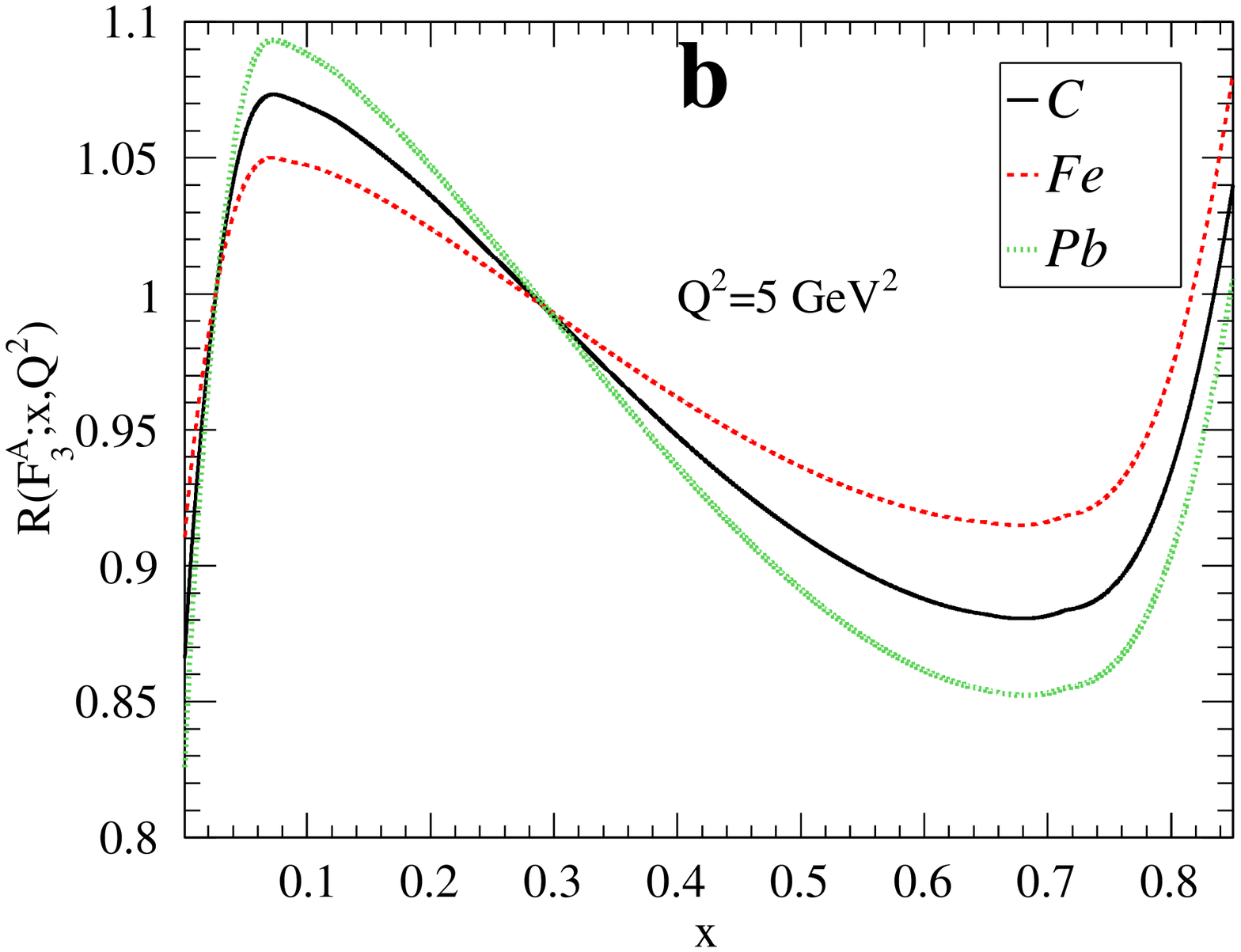}
\centering
\caption{The ratios \textbf{(a)} $R(F_2^A;x,Q^2)$ and \textbf{(b)} $R(F_3^A;x,Q^2)$ as functions of $x$ with $Q^2=5~\textrm{GeV}^2$. The black (solid), red (dotted) and green (dashed) lines correspond to $\rm{C}$, $\rm{Fe}$ and $\rm{Pb}$ respectively.}
\label{R_structure_functions}
\end{figure}

Assuming isospin symmetry between bound neutron and proton, one has
\begin{eqnarray}
f_{u, \bar{u}}^{n / A}&=&f_{d, \bar{d}}^{p / A}\,; \quad f_{d, \bar{d}}^{n / A}=f_{u, \bar{u}}^{p / A}\,; \quad f_{i'}^{n / A}=f_{i'}^{p/ A} \text { for other flavors. }
\end{eqnarray}
Thus the expression of structure function in Eq.(\ref{calcstrucuturefuncs1}) can be further simplified as
\begin{eqnarray}\label{calcstrucuturefuncs2}
  F_2^A  &=& 2x \left( f_u^{p/A}+f_d^{p/A}+f_{\bar u}^{p/A} +f_{\bar d}^{p/A} \right) \,,
\end{eqnarray}
similarly,
\begin{eqnarray}\label{calc_strucuture_funcs3}
  F_3^A &=&  2\left(f_u^{p/A} +f_d^{p/A}-f_{\bar u}^{p/A}-f_{\bar d}^{p/A} \right) \,.
\end{eqnarray}
The NPDF $f_i^{p/A}(x,Q^2)$ can be defined relative to the free proton PDF $f_i^{p}(x,Q^2)$ as \cite{Eskola:2016oht}
\begin{eqnarray}\label{EPPS16R}
  f_{i}^{p / A}\left(x, Q^{2}\right)=R_{i}^{A}\left(x, Q^{2}\right) f_{i}^{p}\left(x, Q^{2}\right) \,,
\end{eqnarray}
here $R_{i}^{A}\left(x, Q^{2}\right)$ is known as nuclear modification factor. The free proton baseline is CT14NLO \cite{Dulat:2015mca}. Fig.\ref{structure_functions} presents the structure functions $F^A_2(x, Q^2)$ and $x F^A_3(x, Q^2)$ in respect of $x$, with $Q^2$ fixed to $5~\textrm{GeV}^2$ which is reachable in many neutrino-nucleus scattering experiments. One can see that the lines in this figure are very close to each other, for a better illustration of EMC effect, the ratios
\begin{eqnarray}\label{RF2andRF3}
  R(F_2^A;x,Q^2) &=& \frac{ f_u^{p/A}(x,Q^2)+f_d^{p/A}(x,Q^2)+f_{\bar u}^{p/A}(x,Q^2) +f_{\bar d}^{p/A}(x,Q^2) }{ f_u^{p}(x,Q^2)+f_d^{p}(x,Q^2)+f_{\bar u}^{p}(x,Q^2) +f_{\bar d}^{p}(x,Q^2)} \,,\nn\\
  R(F_3^A;x,Q^2) &=& \frac{f_u^{p/A}(x,Q^2) +f_d^{p/A}(x,Q^2)-f_{\bar u}^{p/A}(x,Q^2)-f_{\bar d}^{p/A}(x,Q^2) }{f_u^{p}(x,Q^2) +f_d^{p}(x,Q^2)-f_{\bar u}^{p}(x,Q^2)-f_{\bar d}^{p}(x,Q^2)} \,,
\end{eqnarray}
are presented in Fig.\ref{R_structure_functions}.

\section{Modification Factors for $d^2\sigma/dxdy$}
\label{diffcrosssection}
\begin{figure}[htbp]
\begin{minipage}{0.32\columnwidth}
\centerline{\includegraphics[width=1\columnwidth]{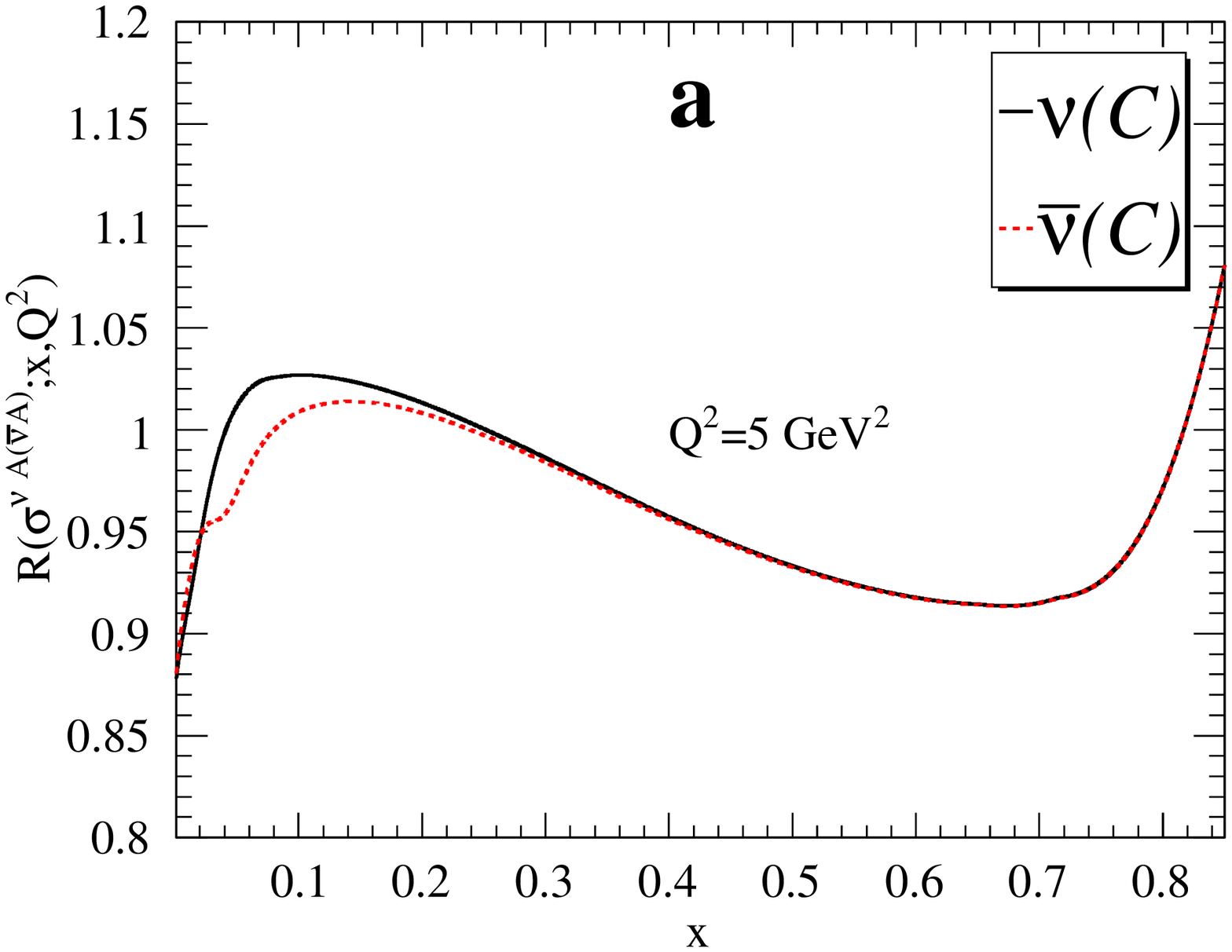}}
\end{minipage}
\begin{minipage}{0.32\columnwidth}
\centering
\centerline{\includegraphics[width=1\columnwidth]{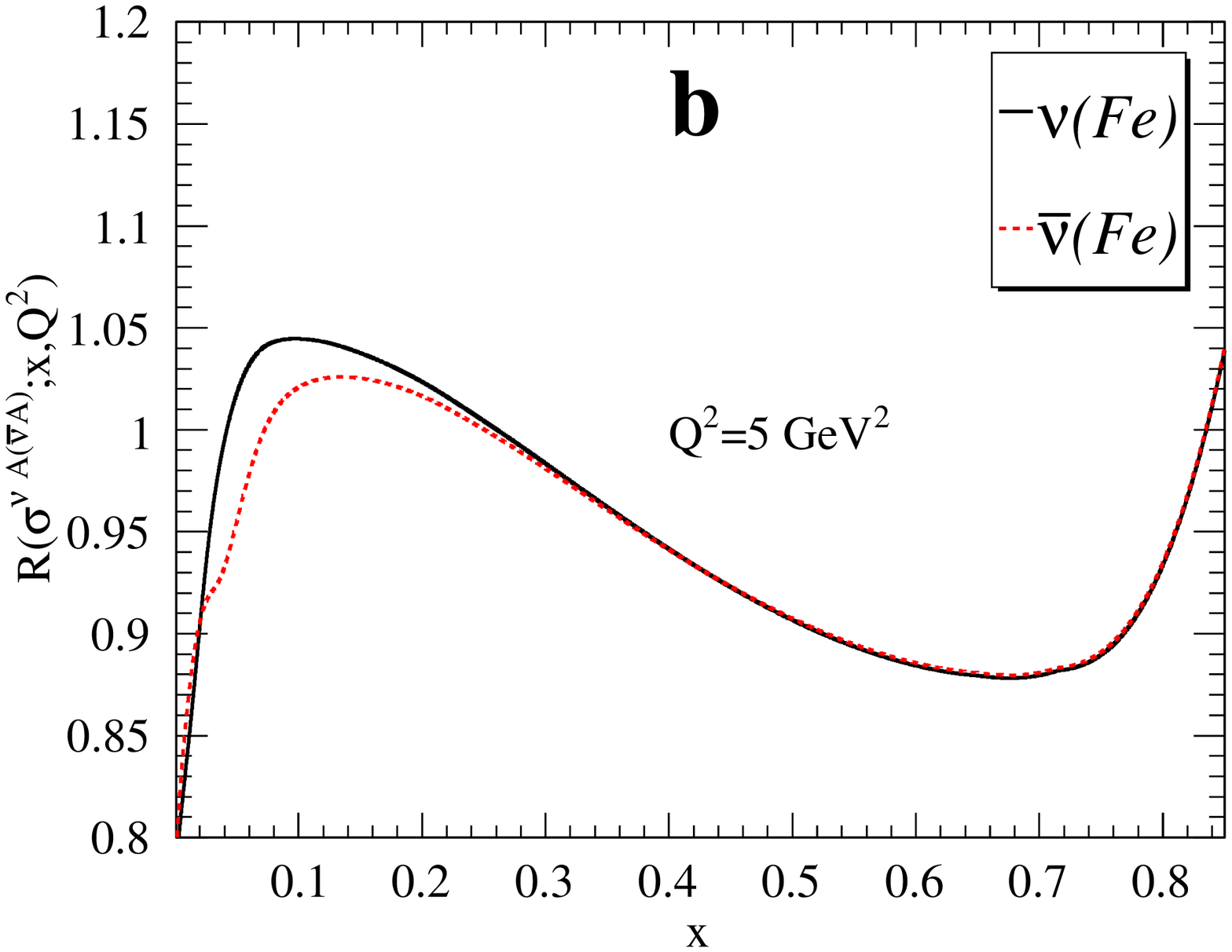}}
\end{minipage}
\begin{minipage}{0.32\columnwidth}
\centering
\centerline{\includegraphics[width=1\columnwidth]{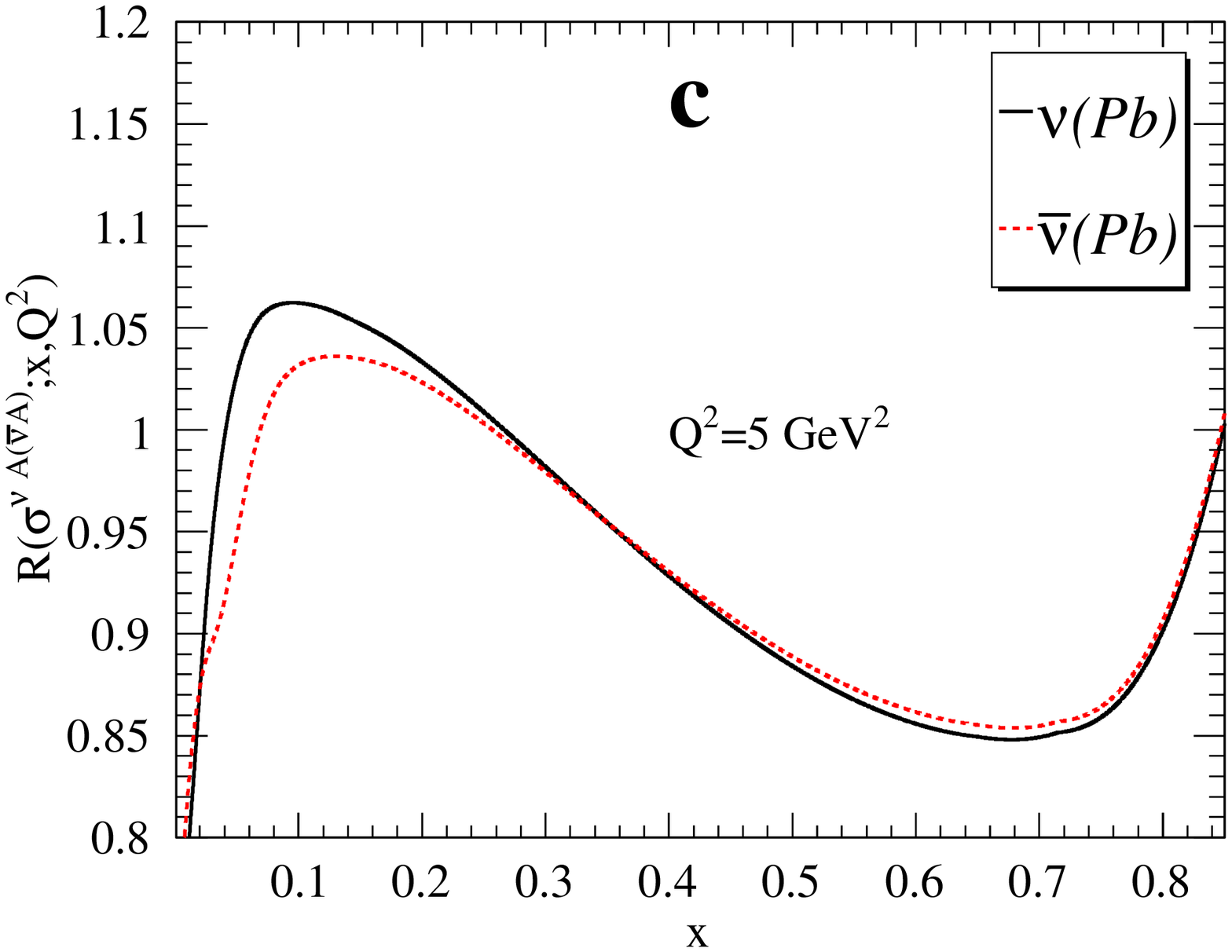}}
\end{minipage}
\caption{Nuclear correction factor $R\left(\sigma^{\nu A(\bar\nu A)}; x,Q^2\right)$ according to Eq.(\ref{comparediff}) in charged-current \textbf{(a)} $\nu(\bar\nu) \textrm{C}$, \textbf{(b)} $\nu(\bar\nu) \textrm{Fe}$ and \textbf{(c)} $\nu(\bar\nu) \textrm{Pb}$ scattering.}
	\label{R_Three_diff_cross}
\end{figure}

Absolute neutrino flux is never known to better than 20-30\%, this makes total cross sections hard to measure accurately, so experimentalists like to measure cross section ratios as well as differential cross section which is one of the least model dependent measurements \cite{Tzanov:2009zz}. In previous section, we have analyzed structure functions of neutrino-nucleus DIS. In this section, we will compare differential cross sections of neutrino-nucleus DIS and neutrino-free nucleon DIS
\begin{eqnarray}\label{comparediff}
  R\left(\sigma^{\nu A}; x,Q^2\right)=\left(\frac{d^2 \sigma}{d x d y}\right)^{\nu A} \Big/ \left(\frac{d^2 \sigma}{d x d y}\right)^{\nu\,\textrm{nucleon}/A} \,,\nn\\
  R\left(\sigma^{\bar\nu A};x,Q^2\right)=\left(\frac{d^2 \sigma}{d x d y}\right)^{\bar\nu A} \Big/ \left(\frac{d^2 \sigma}{d x d y}\right)^{\bar\nu\,\textrm{nucleon}/A} \,,
\end{eqnarray}
here $R\left(\sigma^{\nu A(\bar\nu A)}; x,Q^2\right)$ are called nuclear correction factors. The numerator in Eq.(\ref{comparediff}) can be derived from Eq.(\ref{diff_for_nuandbarnu2}) and for the denominator we have used
\begin{eqnarray}\label{diff_for_nuandbarnu2_free}
\left(\frac{d \sigma}{d x d y}\right)^{\nu\,\textrm{nucleon}/A}&=&\frac{G_{F}^{2} M_{N} E_\nu}{\pi\left(1+Q^{2} / M_{W}^{2}\right)^{2}}\left[\left(1-y+\frac{y^{2}}{2}\right) F_{2}^{\nu\,\textrm{nucleon}/A}+x y\left(1-\frac{y}{2}\right) F_{3}^{\nu\,\textrm{nucleon}/A}\right] \,,\nn\\
\left(\frac{d \sigma}{d x d y}\right)^{\bar{\nu}\,\textrm{nucleon}/A}&=&\frac{G_{F}^{2} M_{N} E_{\bar\nu}}{\pi\left(1+Q^{2} / M_{W}^{2}\right)^{2}}\left[\left(1-y+\frac{y^{2}}{2}\right) F_{2}^{\bar \nu\,\textrm{nucleon}/A}-x y\left(1-\frac{y}{2}\right) F_{3}^{\bar \nu\,\textrm{nucleon}/A}\right] \,,\nn
\end{eqnarray}
with
\begin{eqnarray}
  \left\{\begin{array}{ll}
  F_2^{\nu(\bar\nu)\,\textrm{nucleon}/A}=\frac{Z}{A} F_2^{\nu(\bar\nu) p} + \frac{A-Z}{A} F_{2}^{\nu(\bar\nu) n} \,,\\
  F_3^{\nu(\bar\nu)\,\textrm{nucleon}/A}=\frac{Z}{A} F_3^{\nu(\bar\nu) p} + \frac{A-Z}{A} F_{3}^{\nu(\bar\nu) n} \,.
\end{array} \right. \nn
\end{eqnarray}
Our results are presented in Fig.\ref{R_Three_diff_cross} for $Q^2=5~\textrm{GeV}^2$. The solid lines correspond to neutrino and dashed lines to antineutrino.

\section{Universality of SRC}
\label{universality}
Neutrino-nucleus DIS process is an ideal platform for testing SRC interpretation of the EMC effect. Motivated by the correlation between the EMC effect and the SRC scale factor and according to the universality of the SRC, we parameterize the $u$ and $d$ quark distributions in the EMC region as that for the structure function in Ref.\cite{Frankfurt:1993sp,Segarra:2019gbp} by assuming that all nuclear modifications originate from the nucleon-nucleon SRCs,
\begin{eqnarray}\label{para_about_distributions}
  f_{u_V(d_V)}^{p/A}\left(x, Q^{2}\right)&=&\frac{1}{Z}\left[Z f_{u_V(d_V)}^{p}\left(x, Q^{2}\right)+n_{s r c}^{A} \delta f_{u_V(d_V)}^{p}\left(x, Q^{2}\right)\right]\,,
\end{eqnarray}
where $n^A_{src}$ represents number of $np$ pairs in nucleus $A$, notably the subscript $V$ in $f_{u_V}$ and $f_{d_V}$ means distributions of valence quark since experimental results pointed to the EMC effect being due to a change in the valence quark distributions. $\delta f_{u_V(d_V)}^{p}$ represents the difference between $u(d)$ valence quark distribution in the SRC pair and in the free proton. Substituting Eq.(\ref{para_about_distributions}) into Eq.(\ref{EPPS16R}), we have
\begin{eqnarray}\label{uni_function_of_ud}
  \frac{\delta f_{u_V(d_V)}^{p}\left(x, Q^{2}\right)}{f_{u_V(d_V)}^{p}\left(x, Q^{2}\right)}=\frac{R_{u_V(d_V)}^{A}\left(x, Q^{2}\right)-1}{n_{src}^A/Z_A} \,.
\end{eqnarray}
The left of this equation above is assumed to be nucleus-independent, meaning that the right-hand side should be a universal function which does not depend on the type of nucleus. Therefore, we can parameterize the quark distribution in $A$ in terms of that in $B$,
\begin{eqnarray}\label{use_of_uni_function}
  R_{u_V(d_V)}^{A}\left(x, Q^{2}\right)&=&\frac{n_{s r c}^{A} / Z_{A}}{n_{s r c}^{B} / Z_{B}}\left(R_{u_V(d_V)}^{B}\left(x, Q^{2}\right)-1\right)+1 \nn\\
  &=& \frac{a_2^A}{a_2^B}\frac{Z_{B}}{Z_{A}}\frac{A_A}{A_B}\left(R_{u_V(d_V)}^{B}\left(x, Q^{2}\right)-1\right)+1 \,.
\end{eqnarray}
Here $a_{2}^{A}=\left(n_{s r c}^{A} / A\right) /\left(n_{s r c}^{d} / 2\right)$ is the SRC ratio of nucleus A respect to that of deuteron. This ratio can be measured through the nuclear structure functions at $x>1.5$ region \cite{Weinstein:2010rt,Fomin:2011ng,Arrington:2012ax}.

\begin{figure}[htbp]
\centering
\includegraphics[width=0.45\columnwidth]{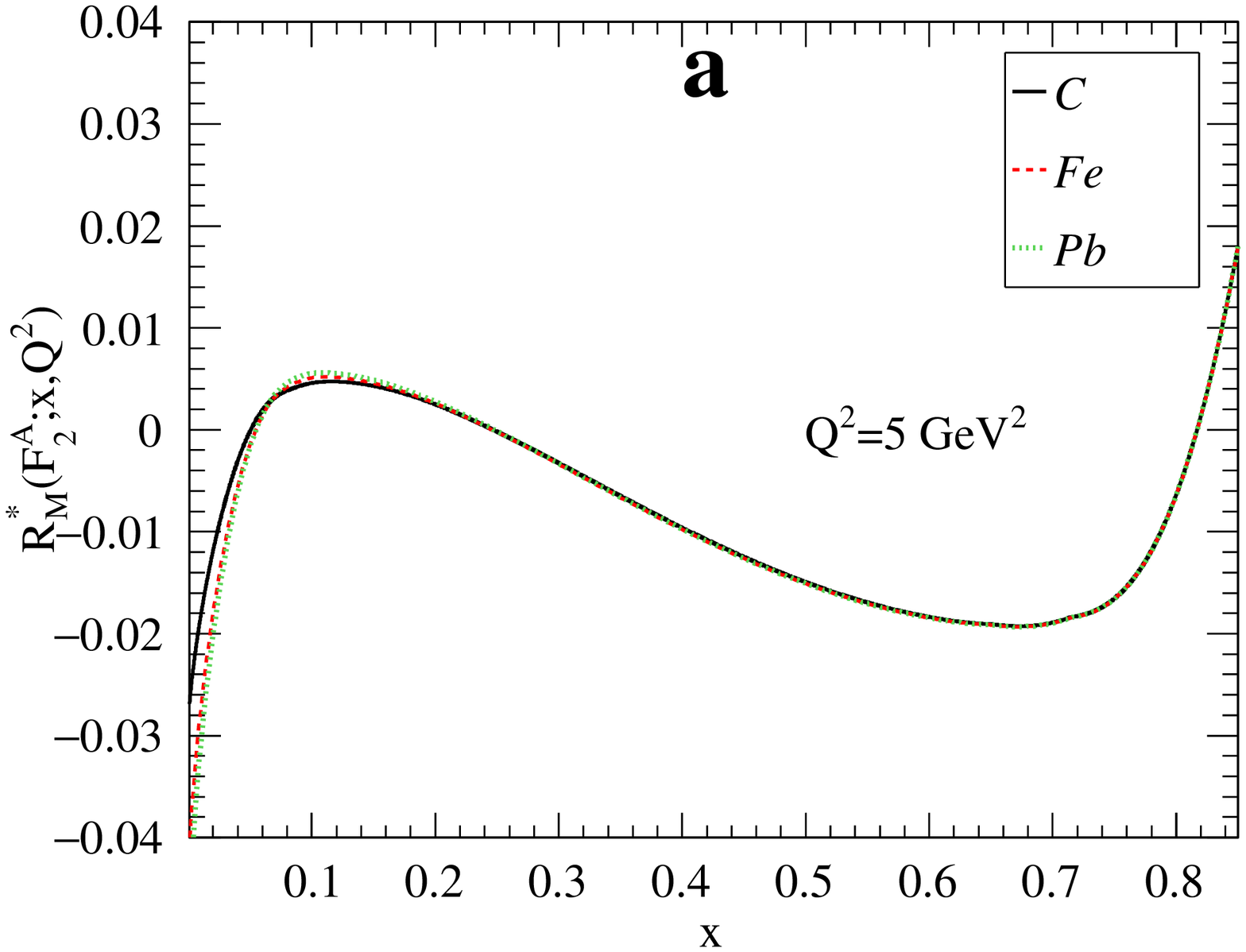}
\centering
\includegraphics[width=0.45\columnwidth]{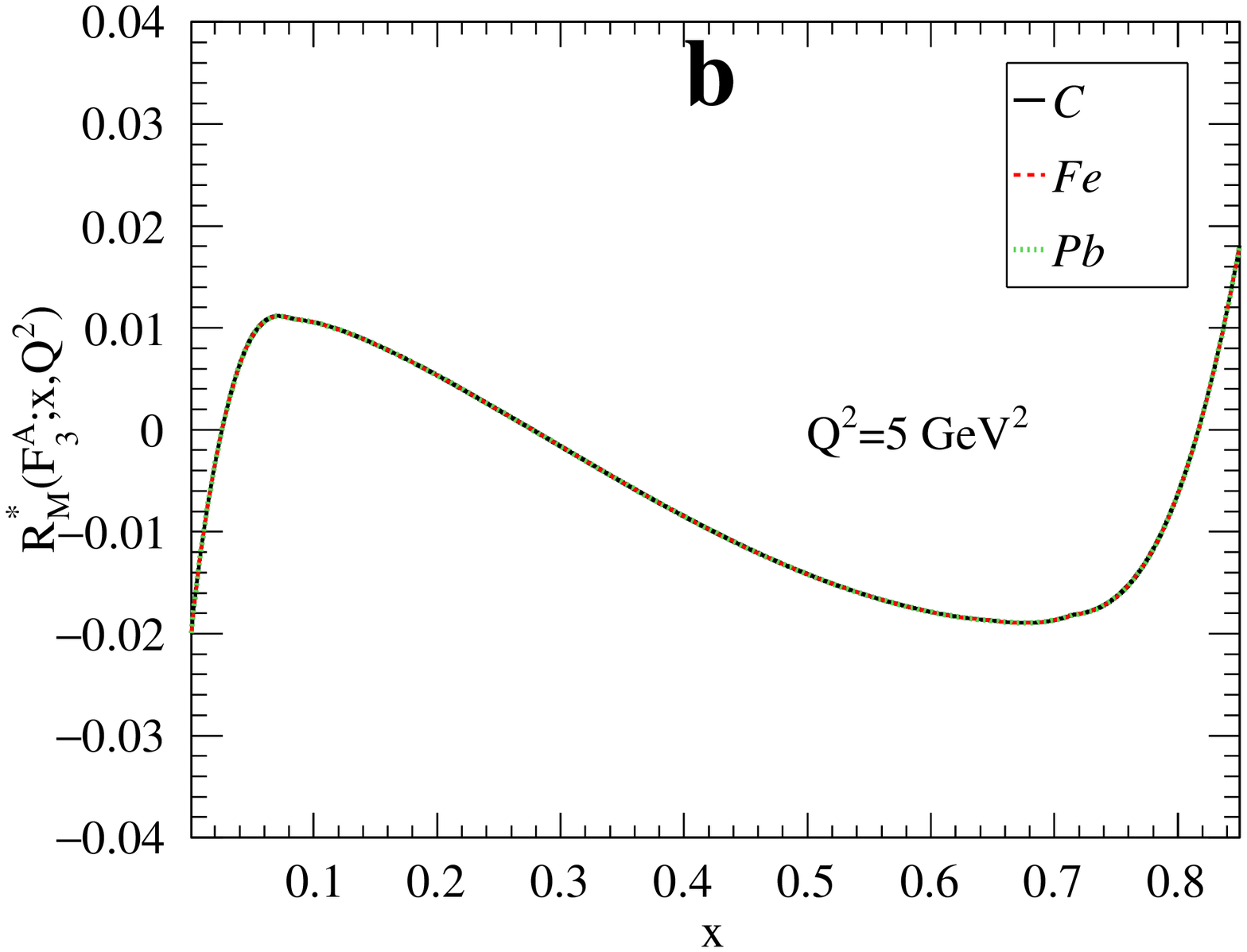}
\centering
\caption{Universality of EMC effects shown in the structure functions with theoretical assumption Eq.(\ref{use_of_uni_function}), here $Q^2=5~\textrm{GeV}^2$. The nucleus-independent (universal) behaviour of the SRC modification is clearly observed.  }
\label{Th_RF23}
\end{figure}

\begin{figure}[htbp]
\centering
\includegraphics[width=0.45\columnwidth]{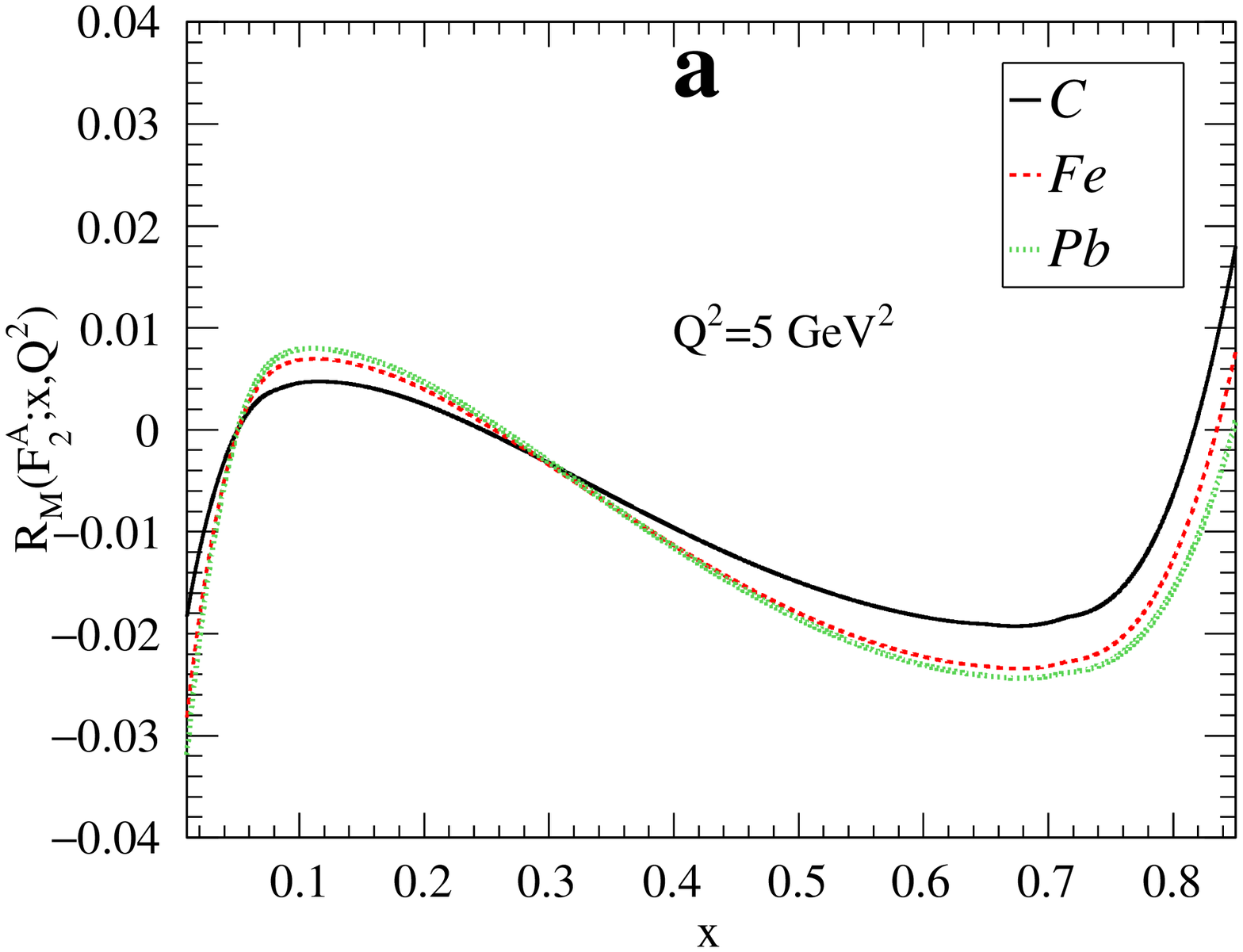}
\centering
\includegraphics[width=0.45\columnwidth]{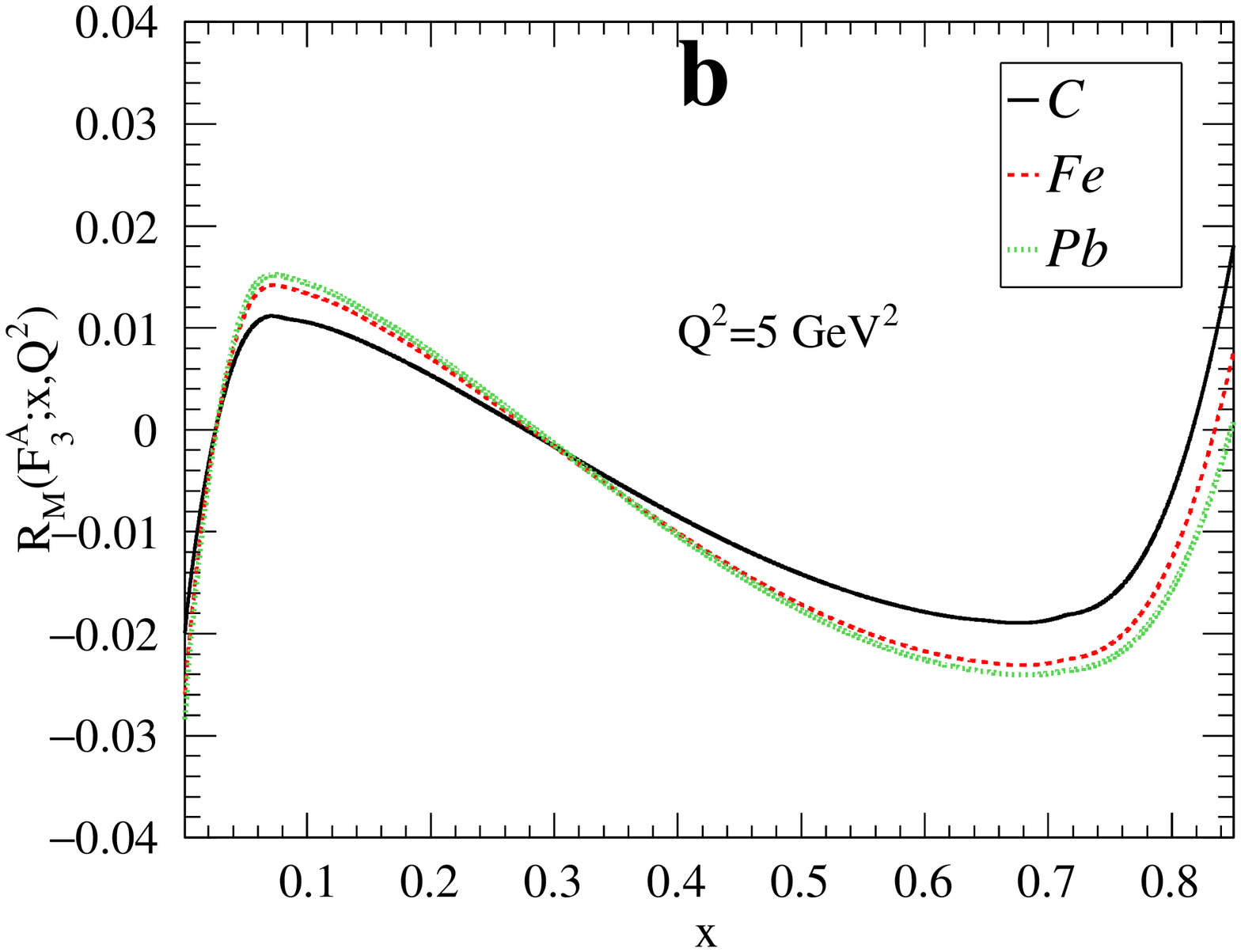}
\centering
\caption{Universality of EMC effects shown in the structure functions with EPPS16 parametrization, here $Q^2=5~\textrm{GeV}^2$.}
\label{Ex_RF23}
\end{figure}

Based on the EPPS16 parametrization of quark NPDFs for carbon and Eq.(\ref{use_of_uni_function}), one can get quark NPDFs for iron and plumbum. With these distributions, the universality of SRC can be illustrated more specifically by introducing one kind of universal modification functions whose functionality is normalizing the ratios defined in Eq.(\ref{RF2andRF3}) by the respective SRC factors
\begin{eqnarray}\label{normalize}
  R^{*}_{M}(F_2^A;x,Q^2)=\frac{2Z_A}{A_A} \frac{R^*(F_2^A;x,Q^2)-1}{a_2^{A}}\,, \qquad R^*_{M}(F_3^A;x,Q^2)=\frac{2Z_A}{A_A} \frac{R^*(F_3^A;x,Q^2)-1}{a_2^{A}} \,,
\end{eqnarray}
here the $R^*$ marked with an asterisk is used to remind that it is derived from Eq.(\ref{use_of_uni_function}), i.e., from the assumption that all nuclear modifications originate from the SRCs in EMC region. $R^*_{M}(F_2^A;x,Q^2)$ and $R^*_{M}(F_3^A;x,Q^2)$ are plotted in Fig.\ref{Th_RF23}, we can clearly see that the three curves of normalized ratios almost shrink into one single line at $0.3<x<0.7$, this is a remarkable prediction from the universality of the SRC contributions. Besides, Fig.\ref{R_structure_functions} already shows the ratios $R(F_2^A;x,Q^2)$ and $R(F_3^A;x,Q^2)$, they can also be normalized by Eq.(\ref{normalize}), denoted as $R_M$. The results of $R_M(F_2^A;x,Q^2)$ and $R_M(F_3^A;x,Q^2)$ are shown in Fig.\ref{Ex_RF23}.

\begin{figure}[htbp]
\begin{minipage}{0.4\columnwidth}
\centerline{\includegraphics[width=1.\columnwidth]{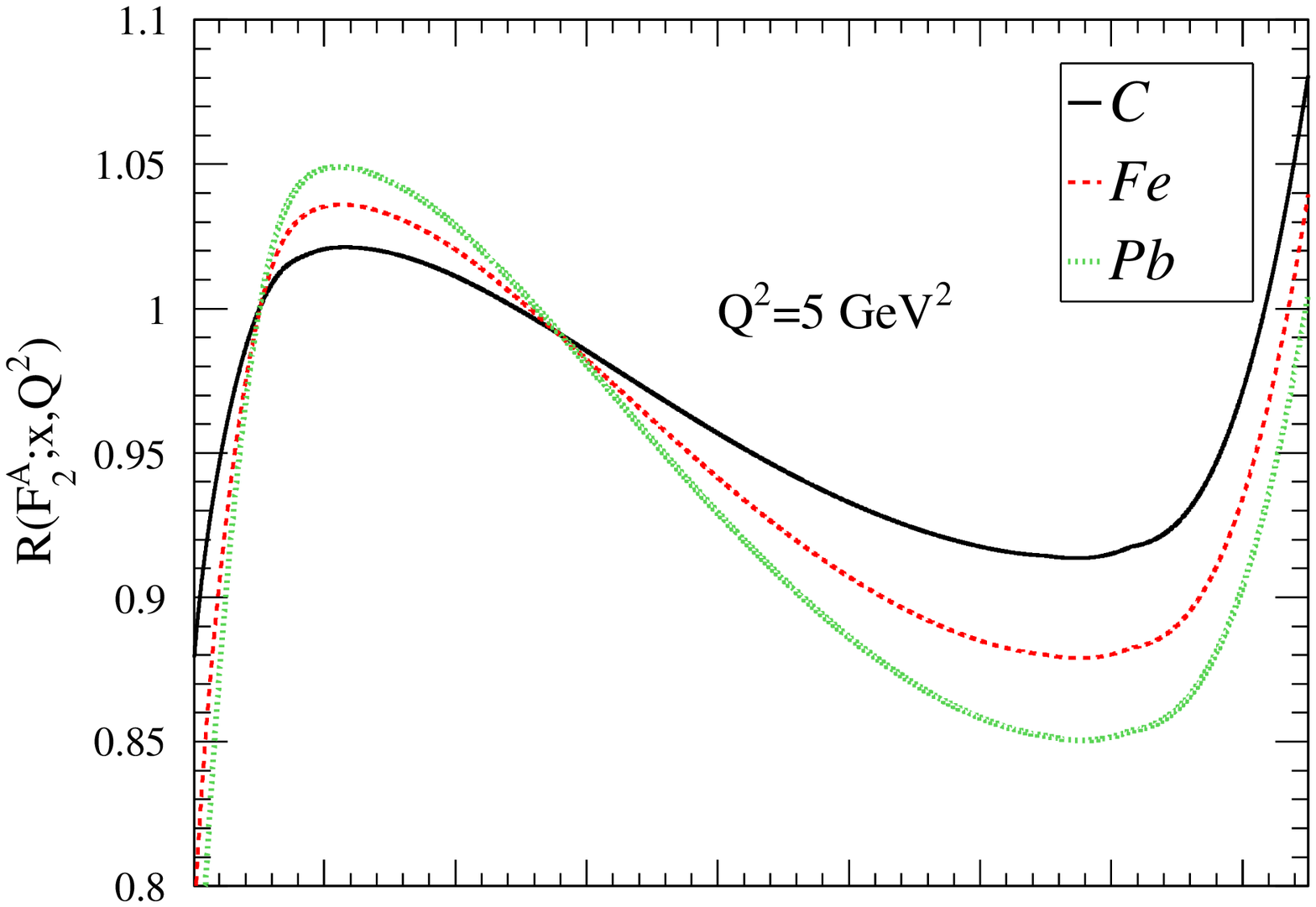}}
\vspace{-18pt}
\centerline{\includegraphics[width=1.\columnwidth]{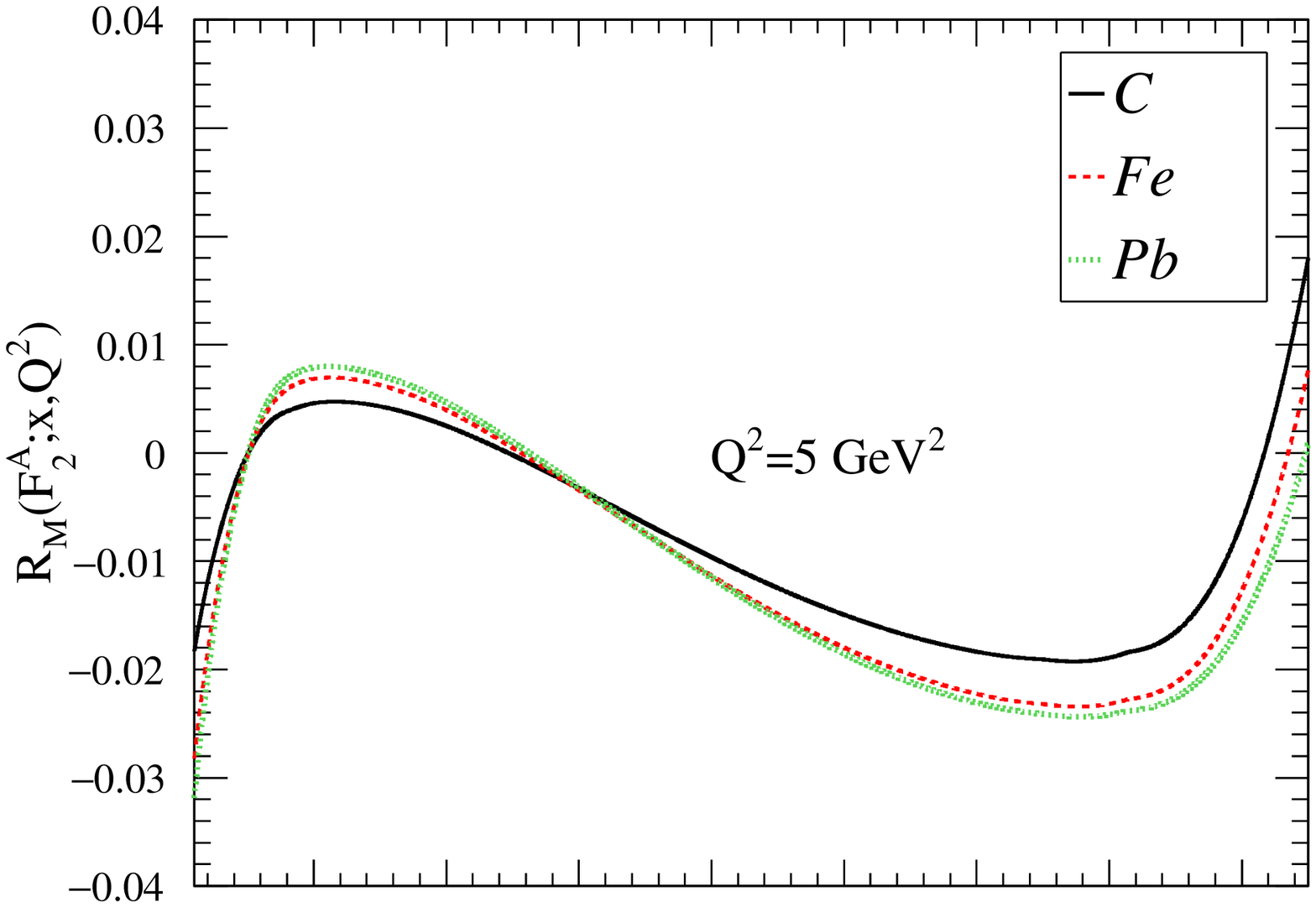}}
\vspace{-18pt}
\centerline{\includegraphics[width=1.\columnwidth]{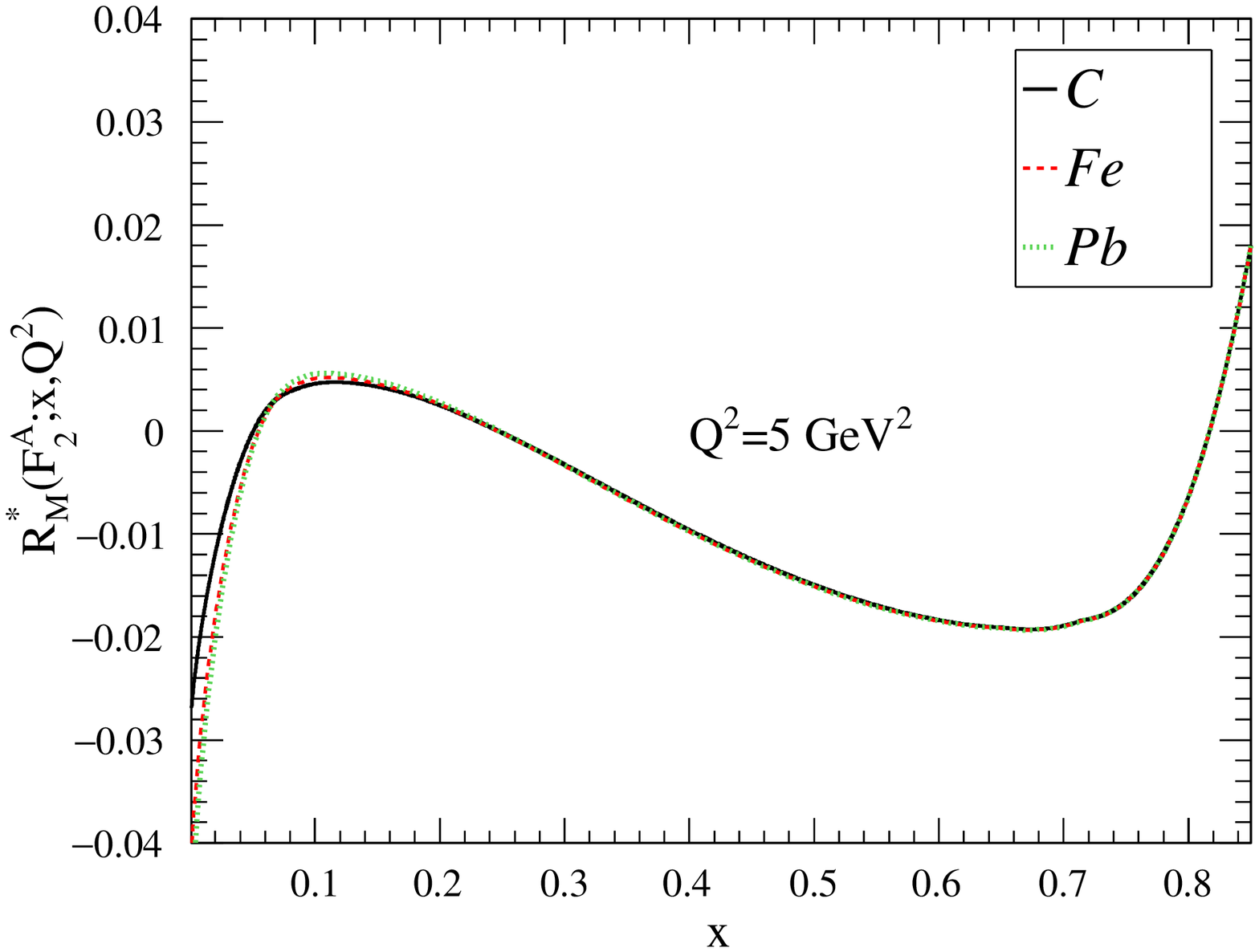}}
\end{minipage}
\begin{minipage}{0.4\columnwidth}
\vspace{-4pt}
\centerline{\includegraphics[width=1.\columnwidth]{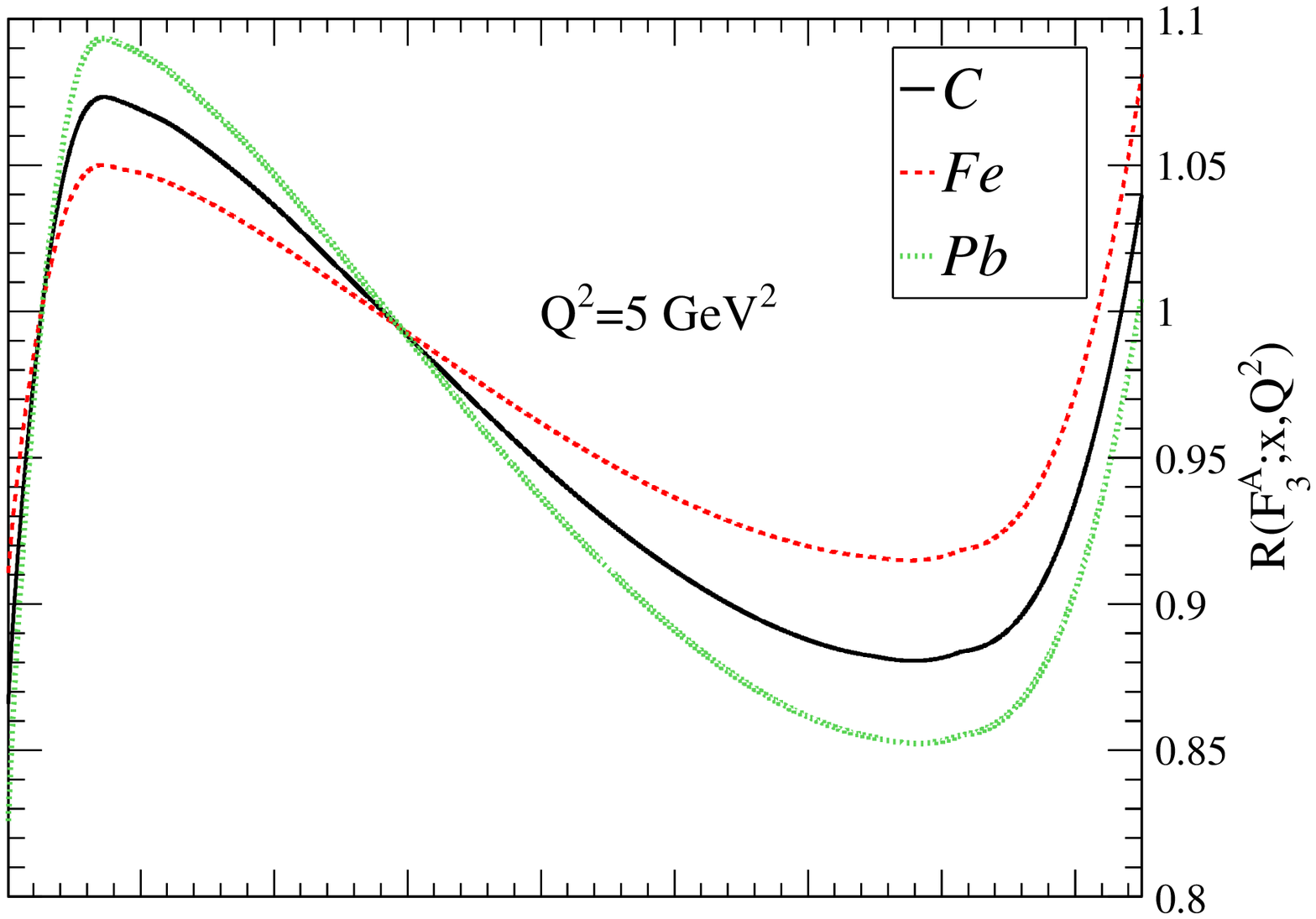}}
\vspace{-18pt}
\centerline{\includegraphics[width=1.\columnwidth]{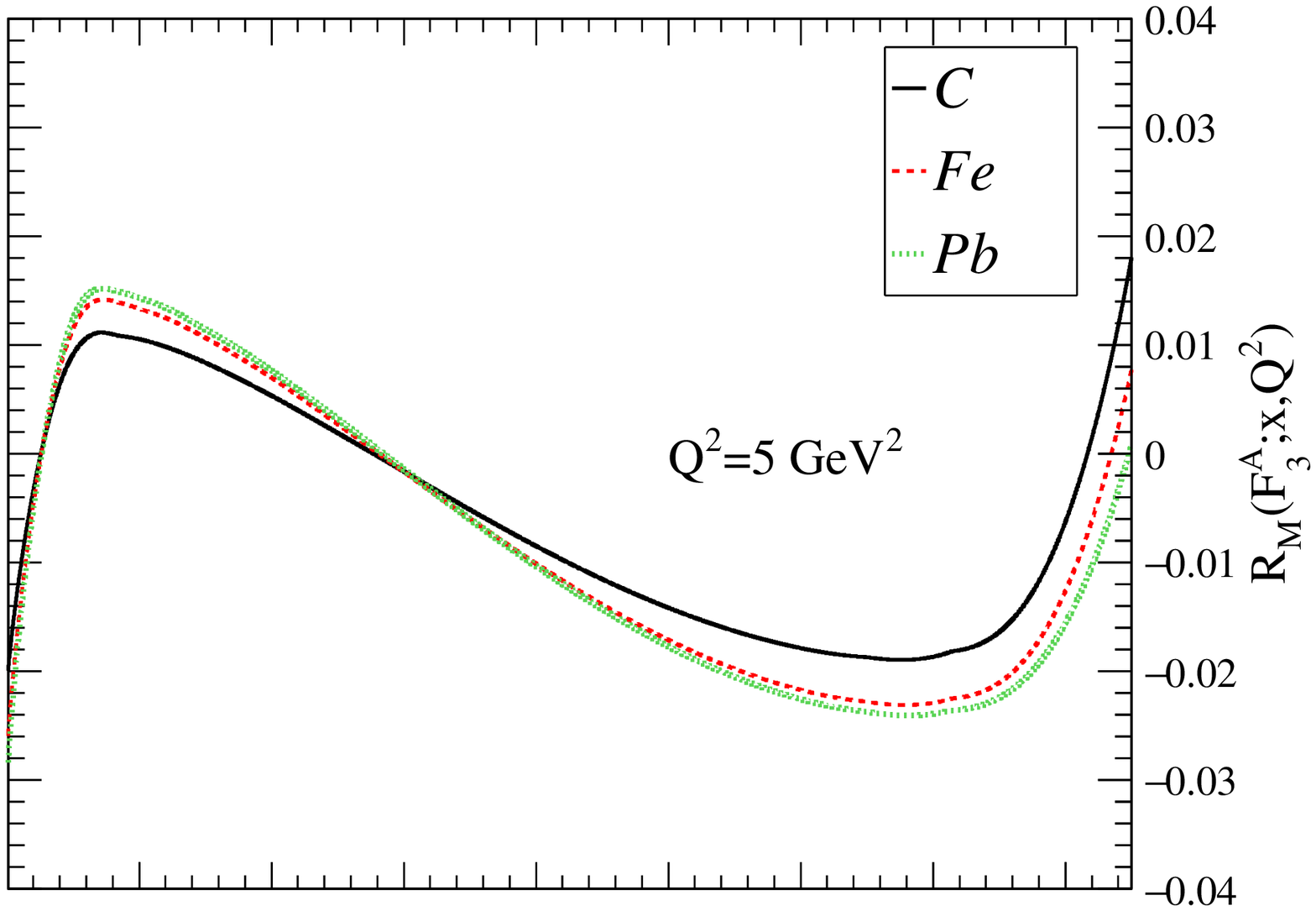}}
\vspace{-18pt}
\centerline{\includegraphics[width=1.\columnwidth]{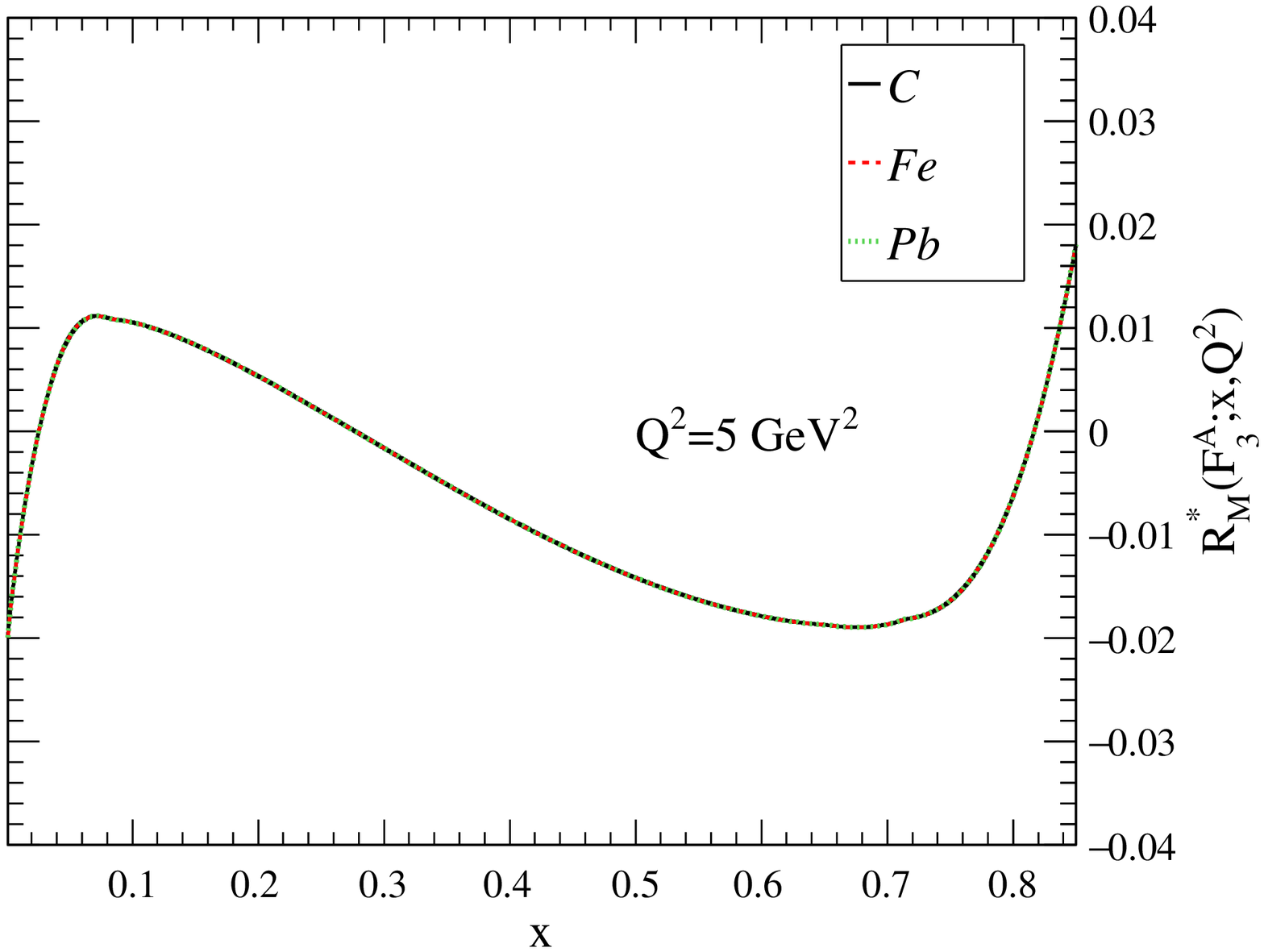}}
\end{minipage}
\caption{Results of $R(F_2^A(F_3^A);x,Q^2)$ (the first row), $R_{M}(F_2^A(F_3^A);x,Q^2)$ (the second row) and $R^*_{M}(F_2^A(F_3^A);x,Q^2)$ (the third row) are presented together in this figure for convenience of comparison.}
\label{combine}
\end{figure}

The predicted  nucleus-independent universal behaviour by the SRC-driven EMC model can be illustrated more clearly if we put these results together. The ratios $R$ defined in Eq.(\ref{RF2andRF3}), the universal modification function $R_M$ with EPPS16 parametrization and universal modification function $R^*_M$ with theoretical assumption have been presented in Fig.\ref{combine}. One can clearly find that after normalized by the respective SRC factors, the ratios of different nucleus tend to shrink substantially, this observation is applicable to both $R_M$ and $R^*_M$. Besides, the shapes of $R_M$ are consistent with $R^*_M$, this consistency supports the theoretical assumption in our paper, indicates EMC effect is mainly caused by SRC pairs in EMC region. Fig.\ref{combine} shows a nontrivial test of SRC universality.

\section{Conclusion}
\label{conclusion}
In summary, we have studied charged-current neutrino-nucleus deep inelastic scattering with particular interests in relationship of SRC and EMC effect. The ratios of structure functions $F^A_2(x,Q^2)$, $x F^A_3(x,Q^2)$ and differential cross sections are presented. We also show the consistency of EMC data and SRC interpretation of EMC effect based on assumption of SRC universality. From these results, one can come to the conclusion the EMC effect in different nuclei can be described by the abundance of SRC pairs and the proposed modification functions are in fact universal. These universal modification functions presented here provide a new clue to understand the fundamental aspects of QCD.

We would like to emphasize that the data used in EPPS16 include not only neutrino-nucleus DIS, but also charged-lepton-nucleus DIS,  Drell-Yan dilepton production and inclusive pion production etc. To purely test the universality of SRC through neutrino-nucleus DIS, more data and new observables were called for. These new data will improve our understanding of nucleon structure substantially.

\section*{Acknowledgements}
The authors are grateful to Professor Wei Wang for inspiring discussions and valuable comments. F.H is supported in part by Natural Science Foundation of China under grant Nos. 11735010, U2032102, 11653003, Natural Science Foundation of Shanghai under grant No. 15DZ2272100, a NSFC-DFG joint grant under grant No. 12061131006 and SCHA 458/22. J.X. is supported in part by National Natural Science Foundation of China under Grant No. 12047545.

\end{document}